\documentclass[twoside,12pt]{article}
\usepackage{indentfirst}
\usepackage{bm}
\usepackage{amsmath, color}
\usepackage{amsfonts}
\usepackage{amssymb}
\usepackage{amsthm}
\usepackage{latexsym}
\newtheorem{thm}{Theorem}

\newtheorem{corollary}{Corollary}
\usepackage{amsmath}

\begin{document}
\renewcommand{\thefootnote}{\fnsymbol{footnote}}
\title{\bf Constructing mutually unbiased bases from unextendible maximally entangled bases
\footnotetext{\\
This work is supported by the National Natural Science Foundation of China under grant Nos. 11101017, 11531004, 11726016 and 11675113,
and Simons Foundation under grant No. 523868, Key Project of Beijing Municipal Commission of Education (KZ201810028042), Beijing Natural Science Foundation (Z190005).}}

\author{\small{Hui Zhao$^1$}\\ \small{$^1$ College of Applied Sciences, Beijing University of Technology, Beijing 100124, China}\\ \small{zhaohui@bjut.edu.cn}\\
\small{Lin Zhang$^1$}\\ \small{$^1$ College of Applied Sciences, Beijing University of Technology, Beijing 100124, China}\\ \small{13963971213@163.com}\\
\small{Shao-Ming Fei$^{2,3}$}\\ \small{$^2$ School of Mathematical Sciences, Capital Normal University, Beijing 100048, China}\\
\small{$^3$ Max-Planck-Institute for Mathematics in the Sciences, 04103 Leipzig, Germany}\\ \small{feishm@mail.cnu.edu.cn}\\
\small{Naihuan Jing$^{4,5}$}\\ \small{$^4$ Department of Mathematics, North Carolina State University, Raleigh, NC 27695, USA}\\
\small{$^5$ Department of Mathematics, Shanghai University, Shanghai 200444, China}\\\small{nathanjing@hotmail.com}
 }
\date{}
\maketitle
%=================== Text begin here =============================================

\vspace*{2mm}

\begin{center}
\begin{minipage}{15.5cm}
\parindent 20pt\footnotesize

We study mutually unbiased bases (MUBs) in which all the bases are unextendible maximally entangled ones. We first present a necessary and sufficient condition of
constructing a pair of MUBs in $C^2 \otimes C^4$. Based on this condition, an analytical and necessary condition for constructing MUBs is given. Moreover we illustrate our approach by some detailed examples in $C^2 \otimes C^4$. The results are generalized to $C^2 \otimes C^d$ $(d\geq 3)$ and a concrete example in $C^2 \otimes C^8$ is given.
\end{minipage}
\end{center}

\begin{center}
\begin{minipage}{15.5cm}
\begin{minipage}[t]{2.3cm}{\bf Keywords:}\end{minipage}
\begin{minipage}[t]{13.1cm}
Mutually unbiased bases, Unextendible maximally entangled bases, unitary matrices
\end{minipage}\par\vglue8pt
{\bf PACS: }03.67.Hk, 03.65.Ud
\end{minipage}
\end{center}
\section{Introduction}
Quantum entanglement is a central resource in quantum information and quantum computation. It is one of the most fascinating features in quantum physics and closely %tightly
related to
some fundamental problems such as estimation of the quantum state.
MUBs can provide an optimal means of determining an ensemble's states given %{\color{red} stated}
by Wootters and Fields [1]. It has good applications such as quantum state tomography [2,3], quantum key distribution [4], quantum teleportation and quantum superdense coding [5,6].
Let ${\ss}_1$=$\{|\phi_i\rangle\}$ and ${\ss}_2$=$\{|\psi_i\rangle\}$, $i=1,2,\cdots,d,$ be two orthonormal bases of a $d$-dimensional complex vector space $C^{d}$, $\langle \phi_i|\phi_i\rangle=\delta_{ij},$ $\langle \psi_i|\psi_i\rangle=\delta_{ij}.$
${\ss}_1$ and ${\ss}_2$ are said to be mutually unbiased if and only if $|\langle\phi_i|\psi_{j}\rangle|=\frac{1}{\sqrt{d}},\forall i,j=1,2,\cdots,d.$ A set of orthonormal bases $\{{\ss}_1,{\ss}_2,\cdots,{\ss}_m$\} in $C^{d}$ is called a set of mutually unbiased bases if every pair of bases in the set is mutually unbiased.

In recent years, there are many constructions of MUBs based on the following bases: product bases(PBs), unextendible product bases (UPBs), maximally entangled bases (MEBs) and unextendible maximally entangled bases (UMEBs). Consider a bipartite quantum system $C^{d}\otimes C^{d'}$ with respective dimension $d$ and $d'$.  A state $|\psi\rangle$ is said to be a $d \otimes d'(d'>d)$ maximally entangled state if and only if for an arbitrary given orthonormal complete basis ${|i_A\rangle}$ of subsystem $A$, there exists an orthonormal complete basis ${|i_B\rangle}$ of subsystem $B$ such that $|\psi\rangle$ can be written as $|\psi\rangle=\frac{1}{\sqrt{d}}\sum^{d-1}_{i=0}{|i_A\rangle}\otimes{|i_B\rangle}$. A method of constructing MUBs based on MEBs in $C^{d}\otimes C^{kd}(k\in Z^{+})$ was given in [7]. Two types of MEBs and their mutually unbiased property in $C^{d}\otimes C^{d'}$ were presented in [8], as detailed examples, some mutually unbiased maximally entangled bases in $C^{2}\otimes C^{4}$, $C^{2}\otimes C^{8}$ and $C^{3}\otimes C^{3}$ were given. Some explicit constructions in $C^{3}\otimes C^{3}$, $C^{4}\otimes C^{4}$, $C^{5}\otimes C^{5}$ and $C^{12}\otimes C^{12}$ were presented in [9]. By using any commutative ring $R$ with $d$ elements and generic character of ($R,+$) instead of $Z_{d}=Z/dZ$, the authors presented some constructions of MUBs [10]. Furthermore, a set of states $\{|\phi_{i}\rangle \in C^{d}\otimes C^{d'}: i=1,2,\cdots,n,n<dd'$\} is called an n-member unextendible maximally entangled basis (UMEB) if and only if: (i)\ $|\phi_{i}\rangle$, $i=1,2,\cdots,n$ are maximally entangled; (ii)\ $\langle\phi_{i}|\phi_{j}\rangle=\delta_{ij}$,$i,j=1,2,\cdots,n$; (iii)\ if $\langle\phi_{i}|\psi\rangle=0$, $\forall$ $ i=1,2,\cdots,n$, then $|\psi\rangle$ cannot be maximally entangled. The constructions of MUBs based on UMEBs in $C^{2}\otimes C^{3}$ and $C^{3}\otimes C^{4}$ were derived in [11-13]. The authors proved that a complete set of mutually unbiased bases of a bipartite system contains a fixed amount of entanglement, independent of the choice of the set [14]. In Ref. [23] a method for constructing the UMEBs has been also presented. Although some considerable progress has been made on MUBs [15-22], there are still some problems that have not been resolved.

In this paper, we study mutually unbiased bases based on unextendible maximally entangled bases. By adding two states to the set of UMEBs from [23], we extend a set of UMEBs to a set of bases over the whole space. Then we focus on how to construct a set of bases that is unbiased with this set of extended bases. In Section 2, we present the approach constructing MUBs in $C^2 \otimes C^4$ and give some explicit examples. A necessary and sufficient condition and an analytically necessary condition of constructing a pair of MUBs in $C^2 \otimes C^4$ are given. In Section 3 we generalize the method to $C^2 \otimes C^d (d\geq3)$ and a pair of MUBs based on UMEBs have been presented in $C^{2}\otimes C^{8}$ as detailed examples. Comments and conclusions are given in Section 4.

\section{MUBs in $C^2 \otimes C^4$} \label{The bound}
Let $\{|0\rangle,|1\rangle$\} and $\{|0'\rangle,|1'\rangle,|2'\rangle,|3'\rangle$\} be the computational basis of $C^2$ and $C^4$, respectively. The following six states constitute an UMEB in $C^2 \otimes C^4$~[23]:
\begin{eqnarray}
|\phi_{n,j}\rangle=\frac{1}{\sqrt{2}}\sum_{a=0}^{1}(-1)^{na}|a\rangle|(j\oplus_{3}a)'\rangle,
\end{eqnarray}
where $j\oplus_{3}a=(j+a)~mod~3$ and $n=0,1, j=0,1,2$.

Adding the following two states to (1),
 \begin{eqnarray}
\nonumber |\phi_{0,3}\rangle=\frac{1}{\sqrt{2}}(|0\rangle+|1\rangle)|3'\rangle,\quad
\nonumber |\phi_{1,3}\rangle=\frac{1}{\sqrt{2}}(|0\rangle-|1\rangle)|3'\rangle,
\end{eqnarray}
then the following eight states constitute a complete orthonormal basis of $C^2 \otimes C^4$:
\begin{equation}
\left\{
\begin{array}{lr}
|\phi_{n,j}\rangle=\frac{1}{\sqrt{2}}\sum_{a=0}^{1}(-1)^{na}|a\rangle|(j\oplus_{3}a)'\rangle,\\
|\phi_{n,3}\rangle=\frac{1}{\sqrt{2}}(|0\rangle+(-1)^{n}|1\rangle)|3'\rangle,
\end{array}
\right.
\end{equation}
where $n=0,1, j=0,1,2$.
Next we construct another complete orthonormal basis based on UMEBs. Let $\{|x_{0}\rangle,|x_{1}\rangle$\} and $\{|x_{0'}\rangle,|x_{1'}\rangle,|x_{2'}\rangle,|x_{3'}\rangle$\} be another basis of $C^2$ and $C^4$, respectively, $S$ and $W$ be the unitary matrices. We can transform the computational basis $\{|0\rangle,|1\rangle$\} and $\{|0'\rangle,|1'\rangle,|2'\rangle,|3'\rangle$\} to $\{|x_{0}\rangle,|x_{1}\rangle$\} and $\{|x_{0'}\rangle,|x_{1'}\rangle,|x_{2'}\rangle,|x_{3'}\rangle$\}, respectively,
\begin{eqnarray}
S(|0\rangle,|1\rangle)=(|x_{0}\rangle,|x_{1}\rangle),~~~~~~~~~~~~~~~~~~~~~~~~~\\
W(|0'\rangle,|1'\rangle,|2'\rangle,|3'\rangle)=(|x_{0'}\rangle,|x_{1'}\rangle,|x_{2'}\rangle,|x_{3'}\rangle).
\end{eqnarray}
According to
\begin{eqnarray}
|\psi_{n,j}\rangle = (S \otimes W)|\phi_{n,j}\rangle,~~~~~~~~~~~~~~~~~~~~~~~
\end{eqnarray}
we have another complete orthonormal basis in  $C^2 \otimes C^4$,
\begin{equation}
\left\{
\begin{array}{lr}
|\psi_{n,j}\rangle=\frac{1}{\sqrt{2}}\sum_{a=0}^{1}(-1)^{na}|x_{a}\rangle|x_{(j\oplus_{3}a)'}\rangle,\\
|\psi_{n,3}\rangle=\frac{1}{\sqrt{2}}(|x_{0}\rangle+(-1)^{n}|x_{1}\rangle)|x_{3'}\rangle,
\end{array}
\right.
\end{equation}
where $n=0,1, j=0,1,2$.
The two bases $\{|\phi_{n,m}\rangle\}$ and $\{|\psi_{n,m}\rangle\}$ in $C^2 \otimes C^4$ are mutually unbiased if and only if they satisfy the following property,
\begin{eqnarray}
|\langle\phi_{n,m}|\psi_{p,q}\rangle|=\frac{1}{2\sqrt{2}}.
\end{eqnarray}
Substituting (5) into (7) we get
\begin{eqnarray}
|\langle\phi_{n,m}|S \otimes W |\phi_{p,q}\rangle|=\frac{1}{2\sqrt{2}},
\end{eqnarray}
where $n,p=0,1;~m,q=0,1,2,3$, $S=(s_{kl})_{2\times2}$, $k,l=1,2$, and $W=(w_{st})_{4\times4}$, $s,t=1,2,3,4$.
Eq.\ (8) implies that the absolute values of the entries of the matrix $S$$\otimes$$W$ under the basis $\{|\phi_{n,m}\rangle$\} have the following forms:
 \begin{eqnarray}
\left[ \begin{array}{cccccccc}
        \frac{1}{2\sqrt{2}}& \cdots& \frac{1}{2\sqrt{2}}&  \frac{1}{2\sqrt{2}}  \\
          \frac{1}{2\sqrt{2}}& \cdots& \frac{1}{2\sqrt{2}}&  \frac{1}{2\sqrt{2}}   \\
        \vdots& \vdots&\vdots&\vdots\\
          \frac{1}{2\sqrt{2}}& \cdots& \frac{1}{2\sqrt{2}}&  \frac{1}{2\sqrt{2}}
           \end{array}
      \right ].
\label{A}
\end{eqnarray}
From (2), we can get an unitary matrix $F$
\begin{eqnarray}
F=
\frac{1}{\sqrt{2}}\left[ \begin{array}{cccccccc}
          1&1&0&0&0&0&0&0   \\
          0&0&1&1&0&0&0&0   \\
          0&0&0&0&1&1&0&0   \\
          0&0&0&0&0&0&1&1     \\
          0&0&0&0&1&-1&0&0    \\
          1&-1&0&0&0&0&0&0   \\
          0&0&1&-1&0&0&0&0   \\
          0&0&0&0&0&0&1&-1
           \end{array}
      \right]
\label{A}
\end{eqnarray}
transforming the computational basis $\{|00'\rangle,|01'\rangle,|02'\rangle,|03'\rangle,|10'\rangle,|11'\rangle,|12'\rangle,|13'\rangle$\} to the basis $\{|\phi_{n,m}\rangle$\}, i.e. $F(|00'\rangle,|01'\rangle,|02'\rangle,|03'\rangle,|10'\rangle,|11'\rangle,|12'\rangle,|13'\rangle)=(|\phi_{0,0}\rangle,|\phi_{0,1}\rangle,\cdots,|\phi_{1,3}\rangle)$.
Therefore the matrix $S\otimes W$ under the basis $\{|\phi_{n,m}\rangle, n=0,1,~m=0,1,2,3$\} is given by
\begin{eqnarray}
F^{\dagger}(S \otimes W)F.
\end{eqnarray}
Where $F^{\dagger}$ denotes Hermitian conjugate of the $F$ matrix. From (7)-(11), we have the following theorem:
\begin{thm}
The two bases $\{|\phi_{n,m}\rangle\}$ and $\{|\psi_{n,m}\rangle\}$ defined as (2) and (6) in $C^2 \otimes C^4$ are mutually unbiased if and only if the following condition is satisfied:
\begin{equation}
\left\{\begin{array}{lr}
|s_{11}w_{k,j}+(-1)^{a}s_{21}w_{k+1,j}+(-1)^{b}s_{12}w_{k,j+1}+(-1)^{c}s_{22}w_{k+1,j+1}|=\frac{1}{\sqrt{2}},\\
|s_{11}w_{k,4}+(-1)^{a}s_{21}w_{k+1,4}+(-1)^{b}s_{12}w_{k,4}+(-1)^{c}s_{22}w_{k+1,4}|=\frac{1}{\sqrt{2}},\\
|s_{11}w_{4,j}+(-1)^{a}s_{21}w_{4,j}+(-1)^{b}s_{12}w_{4,j+1}+(-1)^{c}s_{22}w_{4,j+1}|=\frac{1}{\sqrt{2}},\\
|s_{11}w_{3,j}+(-1)^{a}s_{21}w_{1,j}+(-1)^{b}s_{12}w_{3,j+1}+(-1)^{c}s_{22}w_{1,j+1}|=\frac{1}{\sqrt{2}},\\
|s_{11}w_{k,3}+(-1)^{a}s_{21}w_{k+1,3}+(-1)^{b}s_{12}w_{k,1}+(-1)^{c}s_{22}w_{k+1,1}|=\frac{1}{\sqrt{2}},\\
|s_{11}w_{3,3}+(-1)^{a}s_{21}w_{1,3}+(-1)^{b}s_{12}w_{3,1}+(-1)^{c}s_{22}w_{1,1}|=\frac{1}{\sqrt{2}},\\
|s_{11}w_{3,4}+(-1)^{a}s_{21}w_{1,4}+(-1)^{b}s_{12}w_{3,4}+(-1)^{c}s_{22}w_{1,4}|=\frac{1}{\sqrt{2}},\\
|s_{11}w_{4,3}+(-1)^{a}s_{21}w_{4,3}+(-1)^{b}s_{12}w_{4,1}+(-1)^{c}s_{22}w_{4,1}|=\frac{1}{\sqrt{2}},\\
|s_{11}w_{4,4}+(-1)^{a}s_{21}w_{4,4}+(-1)^{b}s_{12}w_{4,4}+(-1)^{c}s_{22}w_{4,4}|=\frac{1}{\sqrt{2}},
\end{array}
\right.
\end{equation}
where $(a,b,c)\in \{(0,0,0), (0,1,1), (1,0,1), (1,1,0)\}$ and $k,j=1,2$.
\label{1}
\end{thm}
Next we give some examples to illustrate Theorem $1$. Constructing some mutually unbiased bases, for simplification, we can set the matrix S to be a diagonal unitary
matrix. In this case, $|s_{11}|=|s_{22}|=1$ due to the properties of unitary matrix. Therefore one can suppose $s_{11}$=$e^{i\varphi_{1}}$, $s_{22}$=$e^{i\varphi_{2}}$, $w_{st}=r_{st}e^{i\theta_{st}}$, where $s,t=1,2,3,4$ and $i$ is an imaginary unit, i.e. $i^{2}=-1$. Substituting $s_{11}$, $s_{22}$, $w_{11}$, $w_{22}$ and $w_{33}$ into (12), we can get
$\nonumber |r_{11}e^{i(\varphi_{1}+\theta_{11})}\pm r_{22}e^{i(\varphi_{2}+\theta_{22})}|
\nonumber =|r_{22}e^{i(\varphi_{1}+\theta_{22})}\pm r_{33}e^{i(\varphi_{2}+\theta_{33})}|
\nonumber =|r_{33}e^{i(\varphi_{1}+\theta_{33})}\pm r_{11}e^{i(\varphi_{2}+\theta_{11})}|=\frac{1}{\sqrt{2}},
$
therefore $\cos(\varphi_{1}+\theta_{11})\cos(\varphi_{2}+\theta_{22})=\sin(\varphi_{1}+\theta_{11})\sin(\varphi_{2}+\theta_{22})
=\cos(\varphi_{1}+\theta_{22})\cos(\varphi_{2}+\theta_{33})
=\sin(\varphi_{1}+\theta_{22})\sin(\varphi_{2}+\theta_{33})
=\cos(\varphi_{1}+\theta_{33})\cos(\varphi_{2}+\theta_{11})
=\sin(\varphi_{1}+\theta_{33})\sin(\varphi_{2}+\theta_{11})=0.$
So $|r_{11}|=|r_{22}|=|r_{33}|=\frac{1}{2}$. From
$|r_{44}e^{i(\varphi_{1}+\theta_{44})}\pm r_{44}e^{i(\varphi_{2}+\theta_{44})}|=\frac{1}{\sqrt{2}}$ and $W$ is a unitary matrix, we can get
$|r_{44}|=\frac{1}{2}.$ In a similar way, we can get
$|r_{st}|=\frac{1}{2}$ and
\begin{equation}
\left\{
\begin{array}{lr}
|(\varphi_{1}+\theta_{k,j})-(\varphi_{2}+\theta_{k+1,j+1})|=\frac{\pi}{2}+l_{1}\pi,\\
|(\varphi_{1}+\theta_{k,4})-(\varphi_{2}+\theta_{k+1,4})|=\frac{\pi}{2}+l_{2}\pi,\\
|(\varphi_{1}+\theta_{4,j})-(\varphi_{2}+\theta_{4,j+1})|=\frac{\pi}{2}+l_{3}\pi,\\
|(\varphi_{1}+\theta_{3,j})-(\varphi_{2}+\theta_{1,j+1})|=\frac{\pi}{2}+l_{4}\pi,
\end{array}
\right.
\end{equation}
\begin{equation}
\left\{
\nonumber\begin{array}{lr}
|(\varphi_{1}+\theta_{k,3})-(\varphi_{2}+\theta_{k+1,1})|=\frac{\pi}{2}+l_{5}\pi,\\
|(\varphi_{1}+\theta_{33})-(\varphi_{2}+\theta_{11})|=\frac{\pi}{2}+l_{6}\pi,\\
|(\varphi_{1}+\theta_{34})-(\varphi_{2}+\theta_{14})|=\frac{\pi}{2}+l_{7}\pi,\\
|(\varphi_{1}+\theta_{43})-(\varphi_{2}+\theta_{41})|=\frac{\pi}{2}+l_{8}\pi,\\
|(\varphi_{1}+\theta_{44})-(\varphi_{2}+\theta_{44})|=\frac{\pi}{2}+l_{9}\pi,
\end{array}
\right.
\end{equation}
where $k,j=1,2$, $l_{n}\in \{0,1,-1\}$ and $n=1,2,\cdots,9.$
Therefore we obtain a necessary condition of constructing a pair of MUBs based on UMEBs as follows.
\begin{corollary}
If the entries of diagonal unitary matrix $S$ and unitary matrix $W$ satisfying $|r_{st}|=\frac{1}{2}$ and Eq.\ (13), the two bases $\{|\phi_{n,m}\rangle\}$ and $\{|\psi_{n,m}\rangle\}$ defined as (2) and (6) in $C^2 \otimes C^4$ are mutually unbiased.
\label{1}
\end{corollary}
Now we illustrate our construction by some detailed examples according to Corollary $1$.\\
Example 1: According to Corollary $1$, we choose $\varphi_{1}=\frac{\pi}{2},\varphi_{2}=0$, $\theta_{st}=0$ except for $\theta_{11}=\theta_{22}=\theta_{33}=\theta_{44}=\pi$ and $r_{st}=\frac{1}{2}, s,t=1,2,3,4$. Therefore
\begin{eqnarray}
S=
\left[ \begin{array}{cccccccc}
i&0\\
0&1\\
\end{array}
\right ],\quad
 W=
 \frac{1}{2}\left[ \begin{array}{cccccccc}
-1&1&1&1\\
1&-1&1&1\\
1&1&-1&1\\
1&1&1&-1\\
\end{array}
\right ].
\label{A}
\end{eqnarray}
The unitary matrix $S$ can transform the computational basis $\{|0\rangle,|1\rangle\}$ to $\{|x_{0}\rangle,|x_{1}\rangle\}$. From (3), we have
\begin{eqnarray}
\nonumber |x_{0}\rangle=(i,0)^{T},\quad |x_{1}\rangle=(0,1)^{T},
\end{eqnarray}
where $|\phi\rangle^{T}$ denotes the transpose of the state $|\phi\rangle$. According to (4), the unitary matrix $W$ can transform the computational basis $\{|0'\rangle,|1'\rangle,|2'\rangle,|3'\rangle\}$ to $\{|x_{0'}\rangle,|x_{1'}\rangle,|x_{2'}\rangle,|x_{3'}\rangle\}$, then
\begin{eqnarray}
\nonumber |x_{0'}\rangle=\frac{1}{2}(-1,1,1,1)^{T},\quad|x_{1'}\rangle=\frac{1}{2}(1,-1,1,1)^{T},\\
\nonumber |x_{2'}\rangle=\frac{1}{2}(1,1,-1,1)^{T},\quad|x_{3'}\rangle=\frac{1}{2}(1,1,1,-1)^{T}.
\end{eqnarray}
Therefore we obtain the another basis based on UMEBs as follows
\begin{eqnarray}
\nonumber|\psi_{0,0}\rangle=\frac{1}{\sqrt{2}}(|x_{0}\rangle|x_{0'}\rangle+|x_{1}\rangle|x_{1'}\rangle)=\frac{1}{\sqrt{2}}(i|0\rangle|x_{0'}\rangle+|1\rangle|x_{1'}\rangle),\\
\nonumber|\psi_{0,1}\rangle=\frac{1}{\sqrt{2}}(|x_{0}\rangle|x_{1'}\rangle+|x_{1}\rangle|x_{2'}\rangle)=\frac{1}{\sqrt{2}}(i|0\rangle|x_{1'}\rangle+|1\rangle|x_{2'}\rangle),\\
\nonumber|\psi_{0,2}\rangle=\frac{1}{\sqrt{2}}(|x_{0}\rangle|x_{2'}\rangle+|x_{1}\rangle|x_{0'}\rangle)=\frac{1}{\sqrt{2}}(i|0\rangle|x_{2'}\rangle+|1\rangle|x_{0'}\rangle),\\
|\psi_{0,3}\rangle=\frac{1}{\sqrt{2}}(|x_{0}\rangle+|x_{1}\rangle)|x_{3'}\rangle=\frac{1}{\sqrt{2}}(i|0\rangle+|1\rangle)|x_{3'}\rangle,~~~~~~~~~~~\\
\nonumber|\psi_{1,0}\rangle=\frac{1}{\sqrt{2}}(|x_{0}\rangle|x_{0'}\rangle-|x_{1}\rangle|x_{1'}\rangle)=\frac{1}{\sqrt{2}}(i|0\rangle|x_{0'}\rangle-|1\rangle|x_{1'}\rangle),\\
\nonumber|\psi_{1,1}\rangle=\frac{1}{\sqrt{2}}(|x_{0}\rangle|x_{1'}\rangle-|x_{1}\rangle|x_{2'}\rangle)=\frac{1}{\sqrt{2}}(i|0\rangle|x_{1'}\rangle-|1\rangle|x_{2'}\rangle),\\
\nonumber|\psi_{1,2}\rangle=\frac{1}{\sqrt{2}}(|x_{0}\rangle|x_{2'}\rangle-|x_{1}\rangle|x_{0'}\rangle)=\frac{1}{\sqrt{2}}(i|0\rangle|x_{2'}\rangle-|1\rangle|x_{0'}\rangle),\\
\nonumber|\psi_{1,3}\rangle=\frac{1}{\sqrt{2}}(|x_{0}\rangle-|x_{1}\rangle)|x_{3'}\rangle=\frac{1}{\sqrt{2}}(i|0\rangle-|1\rangle)|x_{3'}\rangle,~~~~~~~~~~~
\end{eqnarray}
that is mutually unbiased to the basis given by (2).

\noindent Example 2: We choose $\varphi_{1}=\frac{2\pi}{3},\varphi_{2}=\frac{\pi}{6}$, $r_{st}=\frac{1}{2}, s,t=1,2,3,4$, except for $\theta_{12}=\theta_{23}=\theta_{31}=\theta_{44}=\pi$ all the others $\theta_{st}=0$. Let $\omega_{k}=e^{\frac{k\pi i}{6}},\ k=0,1,\cdots,11$, be the twelfth roots of unity, i.e., $\omega^{12}_{k}=1$. Then
\begin{eqnarray}
S=\left[ \begin{array}{cccccccc}
\omega_{4}&0\\
0&\omega_{1}
\end{array}
\right ],\quad
W=
\frac{1}{2}\left[ \begin{array}{cccccccc}
1&-1&1&1\\
1&1&-1&1\\
-1&1&1&1\\
1&1&1&-1
\end{array}
\right ].
\label{A}
\end{eqnarray}
The unitary matrix $S$ can transform the computational basis $\{|0\rangle,|1\rangle\}$ to $\{|x_{0}\rangle,|x_{1}\rangle\}$, i.e.,
\begin{eqnarray}
\nonumber |x_{0}\rangle=\left(\omega_{4},0\right)^{T},\quad|x_{1}\rangle=\left(0,\omega_{1}\right)^{T}.
\end{eqnarray}
The unitary matrix $W$ can transform the computational bases $\{|0'\rangle,|1'\rangle,|2'\rangle,|3'\rangle\}$ to $\{|x_{0'}\rangle,|x_{1'}\rangle,|x_{2'}\rangle,$ $|x_{3'}\rangle\}$, then
\begin{eqnarray}
\nonumber|x_{0'}\rangle=\frac{1}{2}(1,1,-1,1)^{T},\quad|x_{1'}\rangle=\frac{1}{2}(-1,1,1,1)^{T},\\
\nonumber|x_{2'}\rangle=\frac{1}{2}(1,-1,1,1)^{T},\quad|x_{3'}\rangle=\frac{1}{2}(1,1,1,-1)^{T}.
\end{eqnarray}
Therefore we obtain the another basis based on UMEBs as follows
\begin{eqnarray}
\nonumber|\psi_{0,0}\rangle=\frac{1}{\sqrt{2}}(|x_{0}\rangle|x_{0'}\rangle+|x_{1}\rangle|x_{1'}\rangle)=\frac{1}{\sqrt{2}}\left(\omega_{4}|0\rangle|x_{0'}\rangle+\omega_{1}|1\rangle|x_{1'}\rangle\right),\\
\nonumber|\psi_{0,1}\rangle=\frac{1}{\sqrt{2}}(|x_{0}\rangle|x_{1'}\rangle+|x_{1}\rangle|x_{2'}\rangle)=\frac{1}{\sqrt{2}}\left(\omega_{4}|0\rangle|x_{1'}\rangle+\omega_{1}|1\rangle|x_{2'}\rangle\right),\\
\nonumber|\psi_{0,2}\rangle=\frac{1}{\sqrt{2}}(|x_{0}\rangle|x_{2'}\rangle+|x_{1}\rangle|x_{0'}\rangle)=\frac{1}{\sqrt{2}}\left(\omega_{4}|0\rangle|x_{2'}\rangle+\omega_{1}|1\rangle|x_{0'}\rangle\right),\\
|\psi_{0,3}\rangle=\frac{1}{\sqrt{2}}(|x_{0}\rangle+|x_{1}\rangle)|x_{3'}\rangle=\frac{1}{\sqrt{2}}\left(\omega_{4}|0\rangle+\omega_{1}|1\rangle\right)|x_{3'}\rangle,~~~~~~~~~~~\\
\nonumber|\psi_{1,0}\rangle=\frac{1}{\sqrt{2}}(|x_{0}\rangle|x_{0'}\rangle-|x_{1}\rangle|x_{1'}\rangle)=\frac{1}{\sqrt{2}}\left(\omega_{4}|0\rangle|x_{0'}\rangle-\omega_{1}|1\rangle|x_{1'}\rangle\right),\\
\nonumber|\psi_{1,1}\rangle=\frac{1}{\sqrt{2}}(|x_{0}\rangle|x_{1'}\rangle-|x_{1}\rangle|x_{2'}\rangle)=\frac{1}{\sqrt{2}}\left(\omega_{4}|0\rangle|x_{1'}\rangle-\omega_{1}|1\rangle|x_{2'}\rangle\right),\\
\nonumber|\psi_{1,2}\rangle=\frac{1}{\sqrt{2}}(|x_{0}\rangle|x_{2'}\rangle-|x_{1}\rangle|x_{0'}\rangle)=\frac{1}{\sqrt{2}}\left(\omega_{4}|0\rangle|x_{2'}\rangle-\omega_{1}|1\rangle|x_{0'}\rangle\right),\\
\nonumber|\psi_{1,3}\rangle=\frac{1}{\sqrt{2}}(|x_{0}\rangle-|x_{1}\rangle)|x_{3'}\rangle=\frac{1}{\sqrt{2}}\left(\omega_{4}|0\rangle-\omega_{1}|1\rangle\right)|x_{3'}\rangle,~~~~~~~~~~~
\end{eqnarray}
which is mutually unbiased to the basis given by (2).

\noindent Example 3: By the Corollary $1$, we choose $\varphi_{1}=\frac{\pi}{3},\varphi_{2}=-\frac{\pi}{6}$ and $r_{st}=\frac{1}{2},~s,t=1,2,3,4$, except for $\theta_{23}=\theta_{32}=\theta_{24}=\theta_{34}=\theta_{42}=\theta_{43}=\pi$ all the others $\theta_{st}=0$. We have
\begin{eqnarray}
S=
\left[ \begin{array}{cccccccc}
\omega_{2}&o\\
0&\omega_{11}\\
\end{array}
\right ],\quad
 W=
 \frac{1}{2}\left[ \begin{array}{cccccccc}
1&1&1&1\\
1&1&-1&-1\\
1&-1&1&-1\\
1&-1&-1&1\\
\end{array}
\right ].
\label{A}
\end{eqnarray}
The unitary $S$ can transform the computational basis $\{|0\rangle,|1\rangle\}$ to $\{|x_{0}\rangle,|x_{1}\rangle\}$. From (3), we have
$|x_{0}\rangle=\left(\omega_{2},0\right)^{T},\quad|x_{1}\rangle=\left(0,\omega_{11}\right)^{T}.$
According to (4), the unitary $W$ can transform the computational basis $\{|0'\rangle,|1'\rangle,|2'\rangle,|3'\rangle\}$ to $\{|x_{0'}\rangle,|x_{1'}\rangle,|x_{2'}\rangle,|x_{3'}\rangle\}$, then
\begin{eqnarray}
\nonumber|x_{0'}\rangle=\frac{1}{2}(1,1,1,1)^{T},\quad|x_{1'}\rangle=\frac{1}{2}(1,1,-1,-1)^{T},\\
\nonumber|x_{2'}\rangle=\frac{1}{2}(1,-1,1,-1)^{T},\quad|x_{3'}\rangle=\frac{1}{2}(1,-1,-1,1)^{T}.
\end{eqnarray}
Therefore we obtain the another basis based on UMEBs that is mutually unbiased to the basis given by (2).
\begin{eqnarray}
\nonumber|\psi_{0,0}\rangle=\frac{1}{\sqrt{2}}(|x_{0}\rangle|x_{0'}\rangle+|x_{1}\rangle|x_{1'}\rangle)=\frac{1}{\sqrt{2}}\left(\omega_{2}|0\rangle|x_{0'}\rangle+\omega_{11}|1\rangle|x_{1'}\rangle\right),\\
\nonumber|\psi_{0,1}\rangle=\frac{1}{\sqrt{2}}(|x_{0}\rangle|x_{1'}\rangle+|x_{1}\rangle|x_{2'}\rangle)=\frac{1}{\sqrt{2}}\left(\omega_{2}|0\rangle|x_{1'}\rangle+\omega_{11}|1\rangle|x_{2'}\rangle\right),\\
\nonumber|\psi_{0,2}\rangle=\frac{1}{\sqrt{2}}(|x_{0}\rangle|x_{2'}\rangle+|x_{1}\rangle|x_{0'}\rangle)=\frac{1}{\sqrt{2}}\left(\omega_{2}|0\rangle|x_{2'}\rangle+\omega_{11}|1\rangle|x_{0'}\rangle\right),\\
|\psi_{0,3}\rangle=\frac{1}{\sqrt{2}}(|x_{0}\rangle+|x_{1}\rangle)|x_{3'}\rangle=\frac{1}{\sqrt{2}}\left(\omega_{2}|0\rangle+\omega_{11}|1\rangle\right)|x_{3'}\rangle,~~~~~~~~~~~\\
\nonumber|\psi_{1,0}\rangle=\frac{1}{\sqrt{2}}(|x_{0}\rangle|x_{0'}\rangle-|x_{1}\rangle|x_{1'}\rangle)=\frac{1}{\sqrt{2}}\left(\omega_{2}|0\rangle|x_{0'}\rangle-\omega_{11}|1\rangle|x_{1'}\rangle\right),\\
\nonumber|\psi_{1,1}\rangle=\frac{1}{\sqrt{2}}(|x_{0}\rangle|x_{1'}\rangle-|x_{1}\rangle|x_{2'}\rangle)=\frac{1}{\sqrt{2}}\left(\omega_{2}|0\rangle|x_{1'}\rangle-\omega_{11}|1\rangle|x_{2'}\rangle\right),\\
\nonumber|\psi_{1,2}\rangle=\frac{1}{\sqrt{2}}(|x_{0}\rangle|x_{2'}\rangle-|x_{1}\rangle|x_{0'}\rangle)=\frac{1}{\sqrt{2}}\left(\omega_{2}|0\rangle|x_{2'}\rangle-\omega_{11}|1\rangle|x_{0'}\rangle\right),\\
\nonumber|\psi_{1,3}\rangle=\frac{1}{\sqrt{2}}(|x_{0}\rangle-|x_{1}\rangle)|x_{3'}\rangle=\frac{1}{\sqrt{2}}\left(\omega_{2}|0\rangle-\omega_{11}|1\rangle\right)|x_{3'}\rangle.~~~~~~~~~~~
\end{eqnarray}

\section{MUBs in $C^2 \otimes C^d(d\geq3) $ } \label{The bound}

For the $C^2 \otimes C^4$ quantum systems, one can easily find the unitary matrices $S$ and $W$ to construct a pair of MUBs, and get many detailed examples of MUBs. Generalizing the methods of constructing MUBs to $C^2 \otimes C^d$ quantum systems, one has the difficulties to calculate the unitary matrices $S$ and $W$ for high dimension $d$. We now generalize the method to the case of $C^{2}\otimes C^{d}$. Let $\{|i\rangle$\}$_{i=0}^{1}$ and $\{|j^{'}\rangle$\}$_{j=0}^{d-1}$ denote the computational basis of $C^{2}$ and $C^{d}$ respectively. The following states constitute an UMEB in $C^{2}\otimes C^{d}$ [23]:
\begin{eqnarray}
|\phi_{n,j}\rangle=\frac{1}{\sqrt{2}}\sum_{a=0}^{1}(-1)^{na}|a\rangle|(j\oplus_{d-1}a)'\rangle,\quad n=0,1,\quad j=0,1,\cdots,d-2.
\end{eqnarray}
Adding the following two states to (20)
\begin{eqnarray}
\nonumber |\phi_{n,d-1}\rangle=\frac{1}{\sqrt{2}}(|0\rangle+(-1)^{n}|1\rangle)|(d-1)'\rangle,\quad n=0,1,
\end{eqnarray}
then we get a complete orthonormal basis in $C^{2}\otimes C^{d}$:
\begin{equation}
\left\{
\begin{array}{lr}
|\phi_{n,j}\rangle=\frac{1}{\sqrt{2}}\sum_{a=0}^{1}(-1)^{na}|a\rangle|(j\oplus_{d-1}a)'\rangle,  \\
|\phi_{n,d-1}\rangle=\frac{1}{\sqrt{2}}(|0\rangle+(-1)^{n}|1\rangle)|(d-1)'\rangle,
\end{array}
\right.
\end{equation}
where $n=0,1, j=0,1,\cdots,d-2$.
Let $\{|x_{0}\rangle,|x_{1}\rangle$\} and $\{|x_{j'}\rangle, j'=0,1,\ldots, d-1$\} be another basis of $C^2$ and $C^d$, respectively.
$S=(s_{kl})_{2\times2},k,l=1,2$ and $W=(w_{st})_{d\times d},s,t=1,2,\cdots,d$ be the unitary matrices. We have the following relations
\begin{eqnarray}
S(|0\rangle,|1\rangle)=(|x_{0}\rangle,|x_{1}\rangle),~~~~~~~~~~~~~~~~\\
W(|0'\rangle,|1'\rangle,\ldots,|(d-1)'\rangle)=(|x_{0'}\rangle,|x_{1'}\rangle,\ldots,|x_{(d-1)'}\rangle),\\
|\psi_{n,j}\rangle = (S \otimes W)|\phi_{n,j}\rangle.~~~~~~~~~~~~~~~~~
\end{eqnarray}
Similar to the case of $C^{2}\otimes C^{4}$, we obtain another complete orthonormal basis in $C^2 \otimes C^d$:
\begin{equation}
\left\{
\begin{array}{lr}
|\psi_{n,j}\rangle=\frac{1}{\sqrt{2}}\sum_{a=0}^{1}(-1)^{na}|x_{a}\rangle|x_{(j\oplus_{d-1}a)'}\rangle,\\
|\psi_{n,d-1}\rangle=\frac{1}{\sqrt{2}}(|x_{0}\rangle+(-1)^{n}|x_{1}\rangle)|x_{(d-1)'}\rangle,
\end{array}
\right.
\end{equation}
where $n=0,1, j=0,1,\cdots,d-2$.
The two bases $\{|\phi_{n,m}\rangle$\} and $\{|\psi_{n,m}\rangle$\} in $C^{2}\otimes C^{d}$ are mutually unbiased if and only if they satisfy the following property:
\begin{eqnarray}
|\langle\phi_{n,m}|\psi_{p,q}\rangle|=\frac{1}{\sqrt{2d}},\quad\forall n,p=0,1,\quad m,q=0,1,\cdots,d-1.
\end{eqnarray}
Substituting (24) into (26), we get
\begin{eqnarray}
|\langle\phi_{n,m}|S\otimes W|\phi_{p,q}\rangle|=\frac{1}{\sqrt{2d}}.
\end{eqnarray}
It implies that the absolute values of the entries of the matrix $S$$\otimes$$W$ under the basis $\{|\phi_{n,m}\rangle$\} are all $\frac{1}{\sqrt{2d}}$.
Let $F$ be the unitary matrix transforming the computational basis $|ij'\rangle$ to the basis $\{|\phi_{i,j}\rangle$\}, i.e. $F$($|00'\rangle,|01'\rangle,\cdots,|0(d-1)'\rangle,|10'\rangle,|11'\rangle,\cdots,|1(d-1)'\rangle$)=($|\phi_{0,0}\rangle,|\phi_{0,1}\rangle,\cdots,|\phi_{1,d-1}\rangle$),\\
From (21), we have
 \begin{eqnarray}
F=
\frac{1}{\sqrt{2}}\left[ \begin{array}{ccccccccccccccc}
A\\
B\\
           \end{array}
      \right ],
\label{A}
\end{eqnarray}\\
where
\begin{eqnarray}
A=\left[ \begin{array}{ccccccccccccccc}
1&1&0&0&\cdots&0&0    \\
0&0&1&1&\cdots&0&0    \\
\vdots&\vdots&\vdots&\vdots&\ddots&\vdots&\vdots\\
0&0&0&0&\cdots&1&1\\
\end{array}
      \right ]_{d\times 2d},
B=\left[ \begin{array}{ccccccccccccccc}
 0&0&0&0&\cdots&0&0&1&-1&0&0      \\
1&-1&0&0&\cdots&0&0&0&0&0&0    \\
0&0&1&-1&\cdots&0&0&0&0&0&0    \\
\vdots&\vdots&\vdots&\vdots&\ddots&\vdots&\vdots&\vdots&\vdots&\vdots&\vdots\\
0&0&0&0&\cdots&1&-1&0&0&0&0      \\
0&0&0&0&\cdots&0&0&0&0&1&-1      \\
\end{array}
      \right ]_{d\times 2d}.
\label{A}
\end{eqnarray}
Therefore the matrix $S\otimes W$ under the basis $\{|\phi_{n,m}\rangle$\} are given by
\begin{eqnarray}
F^{\dagger}(S \otimes W)F.
\end{eqnarray}
From (27)-(30), we have the following theorem£º
\begin{thm}
The two bases $\{|\phi_{n,m}\rangle\}$ and $\{|\psi_{n,m}\rangle\}$ defined as (21) and (25) in $C^2 \otimes C^d$ are mutually unbiased if and only if the following condition is satisfied:
\begin{equation}
\left\{\begin{array}{lr}
|s_{11}w_{k,j}+(-1)^{a}s_{21}w_{k+1,j}+(-1)^{b}s_{12}w_{k,j+1}+(-1)^{c}s_{22}w_{k+1,j+1}|=\frac{1}{\sqrt{d}},\\
|s_{11}w_{k,d}+(-1)^{a}s_{21}w_{k+1,d}+(-1)^{b}s_{12}w_{k,d}+(-1)^{c}s_{22}w_{k+1,d}|=\frac{1}{\sqrt{d}},\\
|s_{11}w_{d,j}+(-1)^{a}s_{21}w_{d,j}+(-1)^{b}s_{12}w_{d,j+1}+(-1)^{c}s_{22}w_{d,j+1}|=\frac{1}{\sqrt{d}},\\
|s_{11}w_{d-1,j}+(-1)^{a}s_{21}w_{1,j}+(-1)^{b}s_{12}w_{d-1,j+1}+(-1)^{c}s_{22}w_{1,j+1}|=\frac{1}{\sqrt{d}},\\
|s_{11}w_{k,d-1}+(-1)^{a}s_{21}w_{k+1,d-1}+(-1)^{b}s_{12}w_{k,1}+(-1)^{c}s_{22}w_{k+1,1}|=\frac{1}{\sqrt{d}},\\
|s_{11}w_{d-1,d-1}+(-1)^{a}s_{21}w_{1,d-1}+(-1)^{b}s_{12}w_{d-1,1}+(-1)^{c}s_{22}w_{1,1}|=\frac{1}{\sqrt{d}},\\
|s_{11}w_{d-1,d}+(-1)^{a}s_{21}w_{1,d}+(-1)^{b}s_{12}w_{d-1,d}+(-1)^{c}s_{22}w_{1,d}|=\frac{1}{\sqrt{d}},\\
|s_{11}w_{d,d-1}+(-1)^{a}s_{21}w_{d,d-1}+(-1)^{b}s_{12}w_{d,1}+(-1)^{c}s_{22}w_{d,1}|=\frac{1}{\sqrt{d}},\\
|s_{11}w_{d,d}+(-1)^{a}s_{21}w_{d,d}+(-1)^{b}s_{12}w_{d,d}+(-1)^{c}s_{22}w_{d,d}|=\frac{1}{\sqrt{d}},
\end{array}
\right.
\end{equation}
where $(a,b,c)\in \{(0,0,0), (0,1,1), (1,0,1), (1,1,0)\}$ and $k,j=1,2,\cdots,d-2$.
\label{1}
\end{thm}
Similar to the case of $C^{2}\otimes C^{4}$, for simplification, we set $S$ to be a diagonal unitary matrix and suppose $s_{11}$=$e^{i\varphi_{1}}$, $s_{22}$=$e^{i\varphi_{2}}$ ,
$w_{st}=r_{st}e^{i\theta_{st}},~s,t=1,2,\cdots,d$, where $i$ is an imaginary unit, i.e. $i^{2}=-1$. Substituting them into (31), we get $|r_{st}|=\frac{1}{\sqrt{d}}$ and
\begin{equation}
\left\{
\begin{array}{lr}
|(\varphi_{1}+\theta_{k,j})-(\varphi_{2}+\theta_{k+1,j+1})|=\frac{\pi}{2}+l_{1}\pi,\\
|(\varphi_{1}+\theta_{k,d})-(\varphi_{2}+\theta_{k+1,d})|=\frac{\pi}{2}+l_{2}\pi,\\
|(\varphi_{1}+\theta_{d,j})-(\varphi_{2}+\theta_{d,j+1})|=\frac{\pi}{2}+l_{3}\pi,\\
|(\varphi_{1}+\theta_{k,d-1})-(\varphi_{2}+\theta_{k+1,1})|=\frac{\pi}{2}+l_{4}\pi,\\
|(\varphi_{1}+\theta_{d-1,j})-(\varphi_{2}+\theta_{1,j+1})|=\frac{\pi}{2}+l_{5}\pi,\\
|(\varphi_{1}+\theta_{d-1,d-1})-(\varphi_{2}+\theta_{11})|=\frac{\pi}{2}+l_{6}\pi,\\
|(\varphi_{1}+\theta_{d-1,d})-(\varphi_{2}+\theta_{1,d})|=\frac{\pi}{2}+l_{7}\pi,\\
|(\varphi_{1}+\theta_{d,d-1})-(\varphi_{2}+\theta_{d,1})|=\frac{\pi}{2}+l_{8}\pi,\\
|(\varphi_{1}+\theta_{dd})-(\varphi_{2}+\theta_{dd})|=\frac{\pi}{2}+l_{9}\pi,
\end{array}
\right.
\end{equation}
where $k,j=1,2,\cdots,d-2$, $l_{n}\in \{0,1,-1\}$ and $n=1,2,\cdots,9.$
Therefore we obtain a necessary condition of constructing a pair of MUBs based UMEBs as follows.
\begin{corollary}
If the entries of diagonal unitary matrix $S$ and unitary matrix $W$ satisfying $|r_{st}|=\frac{1}{\sqrt{d}}$ and Eq.\ (32), the two bases $\{|\phi_{n,m}\rangle\}$ and $\{|\psi_{n,m}\rangle\}$ defined as (21) and (25) in $C^2 \otimes C^d$ are mutually unbiased.
\label{1}
\end{corollary}
Now we present a detailed example in $C^2 \otimes C^8$ to illustrate our construction.\\
\noindent Example 4: According to (21), we have the basis
\begin{equation}
\left\{
\begin{array}{lr}
|\phi_{n,j}\rangle=\frac{1}{\sqrt{2}}\sum_{a=0}^{1}(-1)^{na}|a\rangle|(j\oplus_{d-1}a)'\rangle, \\
|\phi_{n,7}\rangle=\frac{1}{\sqrt{2}}(|0\rangle+(-1)^{n}|1\rangle)|7'\rangle,
\end{array}
\right.
\end{equation}
where $n=0,1, j=0,1,\cdots,6$.

Then from Corollary $2$, we choose $\varphi_{1}=\frac{\pi}{2},\varphi_{2}=0$ and $r_{st}=\frac{1}{2\sqrt{2}},~s,t=1,2,\cdots,8$, $\theta_{21}=\theta_{23}=\theta_{25}=\theta_{27}=
\theta_{33}=\theta_{34}=\theta_{37}=\theta_{38}=
\theta_{41}=\theta_{44}=\theta_{45}=\theta_{48}=
\theta_{55}=\theta_{56}=\theta_{57}=\theta_{58}=
\theta_{61}=\theta_{63}=\theta_{66}=\theta_{68}=
\theta_{73}=\theta_{74}=\theta_{75}=\theta_{76}=
\theta_{81}=\theta_{84}=\theta_{86}=\theta_{87}=
\pi$, and the other $\theta=0$. We have
\begin{eqnarray}
S=
\left[ \begin{array}{cccccccccc}
i&0\\
0&1\\
 \end{array}
\right ],\quad
W=
\frac{1}{2\sqrt{2}}\left[ \begin{array}{ccccccccccc}
1&1&1&1&1&1&1&1\\
-1&1&-1&1&-1&1&-1&1\\
1&1&-1&-1&1&1&-1&-1\\
-1&1&1&-1&-1&1&1&-1\\
1&1&1&1&-1&-1&-1&-1\\
-1&1&-1&1&1&-1&1&-1\\
1&1&-1&-1&-1&-1&1&1\\
-1&1&1&-1&1&-1&-1&1
        \end{array}
      \right ].
\label{A}
\end{eqnarray}
The unitary matrix $S$ can transform the computational basis $\{|0\rangle,|1\rangle\}$ to $\{|x_{0}\rangle,|x_{1}\rangle\}$. From (22), we have
\begin{eqnarray}
\nonumber|x_{0}\rangle=(i,0)^{T},\quad|x_{1}\rangle=(0,1)^{T}.
\end{eqnarray}
According to (23), the unitary matrix $W$ can transform the computational basis $\{|0'\rangle,|1'\rangle,\cdots,|7'\rangle\}$ to $\{|x_{0'}\rangle,|x_{1'}\rangle,\cdots,|x_{7'}\rangle\}$, then
\begin{eqnarray}
\nonumber|x_{0'}\rangle=\frac{1}{2\sqrt{2}}(1,-1,1,-1,1,-1,1,-1)^{T},\quad|x_{1'}\rangle=\frac{1}{2\sqrt{2}}(1,1,1,1,1,1,1,1)^{T},\quad\quad\quad\\
\nonumber|x_{2'}\rangle=\frac{1}{2\sqrt{2}}(1,-1,-1,1,1,-1,-1,1)^{T},\quad|x_{3'}\rangle=\frac{1}{2\sqrt{2}}(1,1,-1,-1,1,1,-1,-1)^{T},\\
\nonumber|x_{4'}\rangle=\frac{1}{2\sqrt{2}}(1,-1,1,-1,-1,1,-1,1)^{T},\quad|x_{5'}\rangle=\frac{1}{2\sqrt{2}}(1,1,1,1,-1,-1,-1,-1)^{T},\\
\nonumber|x_{6'}\rangle=\frac{1}{2\sqrt{2}}(1,-1,-1,1,-1,1,1,-1)^{T},\quad|x_{7'}\rangle=\frac{1}{2\sqrt{2}}(1,1,-1,-1,-1,-1,1,1)^{T}.
\end{eqnarray}
Therefore we obtain another basis based on UMEBs which is mutually unbiased to the basis given by (33) as follows
\begin{eqnarray}
\nonumber|\psi_{0,0}\rangle=\frac{1}{\sqrt{2}}(|x_{0}\rangle|x_{0'}\rangle+|x_{1}\rangle|x_{1'}\rangle)=\frac{1}{\sqrt{2}}(i|0\rangle|x_{0'}\rangle+|1\rangle|x_{1'}\rangle),\\
\nonumber|\psi_{0,1}\rangle=\frac{1}{\sqrt{2}}(|x_{0}\rangle|x_{1'}\rangle+|x_{1}\rangle|x_{2'}\rangle)=\frac{1}{\sqrt{2}}(i|0\rangle|x_{1'}\rangle+|1\rangle|x_{2'}\rangle),
\end{eqnarray}
\begin{eqnarray}
\nonumber|\psi_{0,2}\rangle=\frac{1}{\sqrt{2}}(|x_{0}\rangle|x_{2'}\rangle+|x_{1}\rangle|x_{3'}\rangle)=\frac{1}{\sqrt{2}}(i|0\rangle|x_{2'}\rangle+|1\rangle|x_{3'}\rangle),\\
\nonumber|\psi_{0,3}\rangle=\frac{1}{\sqrt{2}}(|x_{0}\rangle|x_{3'}\rangle+|x_{1}\rangle|x_{4'}\rangle)=\frac{1}{\sqrt{2}}(i|0\rangle|x_{3'}\rangle+|1\rangle|x_{4'}\rangle),\\
\nonumber|\psi_{0,4}\rangle=\frac{1}{\sqrt{2}}(|x_{0}\rangle|x_{4'}\rangle+|x_{1}\rangle|x_{5'}\rangle)=\frac{1}{\sqrt{2}}(i|0\rangle|x_{4'}\rangle+|1\rangle|x_{5'}\rangle),\\
\nonumber|\psi_{0,5}\rangle=\frac{1}{\sqrt{2}}(|x_{0}\rangle|x_{5'}\rangle+|x_{1}\rangle|x_{6'}\rangle)=\frac{1}{\sqrt{2}}(i|0\rangle|x_{5'}\rangle+|1\rangle|x_{6'}\rangle),\\
\nonumber|\psi_{0,6}\rangle=\frac{1}{\sqrt{2}}(|x_{0}\rangle|x_{6'}\rangle+|x_{1}\rangle|x_{0'}\rangle)=\frac{1}{\sqrt{2}}(i|0\rangle|x_{6'}\rangle+|1\rangle|x_{0'}\rangle),\\
|\psi_{0,7}\rangle=\frac{1}{\sqrt{2}}(|x_{0}\rangle+|x_{1}\rangle)|x_{7'}\rangle=\frac{1}{\sqrt{2}}(i|0\rangle+|1\rangle)|x_{7'}\rangle,~~~~~~~~~~~\\
\nonumber|\psi_{1,0}\rangle=\frac{1}{\sqrt{2}}(|x_{0}\rangle|x_{0'}\rangle-|x_{1}\rangle|x_{1'}\rangle)=\frac{1}{\sqrt{2}}(i|0\rangle|x_{0'}\rangle-|1\rangle|x_{1'}\rangle),\\
\nonumber|\psi_{1,1}\rangle=\frac{1}{\sqrt{2}}(|x_{0}\rangle|x_{1'}\rangle-|x_{1}\rangle|x_{2'}\rangle)=\frac{1}{\sqrt{2}}(i|0\rangle|x_{1'}\rangle-|1\rangle|x_{2'}\rangle),\\
\nonumber|\psi_{1,2}\rangle=\frac{1}{\sqrt{2}}(|x_{0}\rangle|x_{2'}\rangle-|x_{1}\rangle|x_{3'}\rangle)=\frac{1}{\sqrt{2}}(i|0\rangle|x_{2'}\rangle-|1\rangle|x_{3'}\rangle),\\
\nonumber|\psi_{1,3}\rangle=\frac{1}{\sqrt{2}}(|x_{0}\rangle|x_{3'}\rangle-|x_{1}\rangle|x_{4'}\rangle)=\frac{1}{\sqrt{2}}(i|0\rangle|x_{3'}\rangle-|1\rangle|x_{4'}\rangle),\\
\nonumber|\psi_{1,4}\rangle=\frac{1}{\sqrt{2}}(|x_{0}\rangle|x_{4'}\rangle-|x_{1}\rangle|x_{5'}\rangle)=\frac{1}{\sqrt{2}}(i|0\rangle|x_{4'}\rangle-|1\rangle|x_{5'}\rangle),\\
\nonumber|\psi_{1,5}\rangle=\frac{1}{\sqrt{2}}(|x_{0}\rangle|x_{5'}\rangle-|x_{1}\rangle|x_{6'}\rangle)=\frac{1}{\sqrt{2}}(i|0\rangle|x_{5'}\rangle-|1\rangle|x_{6'}\rangle),\\
\nonumber|\psi_{1,6}\rangle=\frac{1}{\sqrt{2}}(|x_{0}\rangle|x_{6'}\rangle-|x_{1}\rangle|x_{0'}\rangle)=\frac{1}{\sqrt{2}}(i|0\rangle|x_{6'}\rangle-|1\rangle|x_{0'}\rangle),\\
\nonumber|\psi_{1,7}\rangle=\frac{1}{\sqrt{2}}(|x_{0}\rangle-|x_{1}\rangle)|x_{7'}\rangle=\frac{1}{\sqrt{2}}(i|0\rangle-|1\rangle)|x_{7'}\rangle.~~~~~~~~~~~
\end{eqnarray}
\textbf {Remark:} In [7], the authors constructed MUBs comprised of only maximally entangled bases in $ C^d \otimes C^{kd}, k\in Z^+$. However we construct the bases in $ C^2 \otimes C^{d}$ by adding $|\phi_{n,d-1}\rangle=\frac{1}{\sqrt{2}}(|0\rangle+(-1)^{n}|1\rangle)|(d-1)'\rangle~,n=0,1$ to a set of UMEBs. Hence our MUBs are not maximally entangled. So our construction method and the MUBs obtained from it are completely different from the Ref.\ [7]. In addition, when the dimension of the first quantum system is 2, MUBs can only be constructed
for $ C^2 \otimes C^{2d}$ in [7]. But using our method, MUBs can be constructed for the dimension of the second quantum system, no matter odd or even.
\section{ Conclusion  }
We have studied the MUBs based on UMEBs. By using unitary matrices $S$ and $W$, we present a necessary and sufficient condition of constructing a pair of MUBs in $C^{2}\otimes C^{4}$. An analytical necessary condition for constructing MUBs has been given by choosing $S$ to be diagonal unitary matrix and $W$ be unitary matrix. Some explicit examples are given to illustrate our approach. Then we generalize the method to $C^{2}\otimes C^{d}$ for $d\geq3$, a necessary and sufficient condition and a necessary condition of constructing a pair of MUBs in $C^{2}\otimes C^{d}$ are given. As a detailed example, a pair of MUBs based on UMEBs has been presented in $C^{2}\otimes C^{8}$. These results can help constructing MUBs. It would be interesting to investigate some applications in quantum information by these MUBs.

\end{document}